\documentclass[conference]{IEEEtran}
\IEEEoverridecommandlockouts
\usepackage{cite}
\usepackage{amsmath,amssymb,amsfonts}
\usepackage{algorithmic}
\usepackage{graphicx}
\usepackage{textcomp}
\usepackage{xcolor}

\usepackage[ruled]{algorithm2e}
\usepackage{multirow}
\usepackage{cases}

\usepackage{subfloat}
\usepackage{subfig}
\usepackage{caption}
\usepackage{booktabs}
\usepackage{flushend}

\def\BibTeX{{\rm B\kern-.05em{\sc i\kern-.025em b}\kern-.08em
    T\kern-.1667em\lower.7ex\hbox{E}\kern-.125emX}}
\begin{document}

\title{IOTA Tangle 2.0: Toward a Scalable, Decentralized, Smart, and Autonomous IoT Ecosystem}

\author{
\IEEEauthorblockN{Nathan Sealey, Adnan Aijaz, and Ben Holden}
\IEEEauthorblockA{
\text{Bristol Research and Innovation Laboratory, Toshiba Europe Ltd., Bristol, United Kingdom}\\
firstname.lastname@toshiba-bril.com}
}    

\maketitle

\begin{abstract}
IOTA Tangle is a distributed ledger technology (DLT), primarily designed for Internet-of-Things (IoT) networks and applications. IOTA Tangle utilizes a direct acyclic graph (DAG) structure for the ledger, with its protocol offering features attractive to the IoT domain, over most blockchain alternatives, such as feeless transactions, higher achievable transactions per second (TPS), and lower energy consumption. The original IOTA implementation relied on a bootstrap centralized coordinator solution for consensus which limited its degree of decentralization and scalability. This concern, alongside other limitations to its adoption, such as lack of smart contracts, are being addressed with the release of IOTA 2.0. This update brings with it significant changes in order to remove the coordinator and achieve a scalable decentralized solution. To this end, this paper provides a technical overview of the key features of IOTA 2.0 while discussing their relevance and benefits for the wider IoT ecosystem. The paper also provides performance insights and future research directions for IOTA 2.0. 
\end{abstract}


\section{Introduction}
%
%
%
%

Distributed ledger technology is promising to address the  issues originating from existing centralized Internet-of-Things (IoT) architectures. The key features of DLT which are beneficial to IoT implementations include decentralization, immutability, auditability, and cryptographic security. Decentralization allows network-wide information exchange via peer-to-peer (P2P) communication, removing the need for validation or control by an authorized and trusted central third party. This can reduce system costs as maintenance and running of said centralized main sever is no longer required and has the additional benefit to users that they no longer must rely on and trust any centralized third party. Immutability prevents tampering of data and facilitates full history tracking of all data exchanges. Auditability is important for IoT networks with each device having the option to store a full copy of the ledger, if required, and timestamped transaction records ensure all transactions and data exchanges are verifiable. \textcolor{black}{Each DLT utilizes a unique cryptographic algorithm to secure its protocol which equips IoT networks with resilience against adversarial attacks. Some of the employed cryptographic algorithms, such as those that harness one-time signature schemes, are quantum-secure, whilst those that use RSA or elliptic curve encryption are susceptible to attacks from quantum computers.}


Blockchain is the most well-known form of DLT; however, associated fees and energy concerns of the mining process as well as limited transaction throughput make many blockchain solutions unsuitable for IoT applications. DLTs utilizing directed acyclic graph (DAG) topology do away with the singular block chains, thereby storing transactions directly within the ledger as opposed to within blocks. The DAG structure allows asynchronous parallel attachment of transactions therefore facilitating a larger throughput and addressing the scalability concerns of blockchain. DAG-based DLTs also do not use mining, with their protocol consequently facilitating feeless transactions and much more lightweight devices due to the reduced energy demand.

\begin{figure}
    \centering
      \subfloat[\label{dag}]{%
       \centering
    \includegraphics[width=\linewidth]{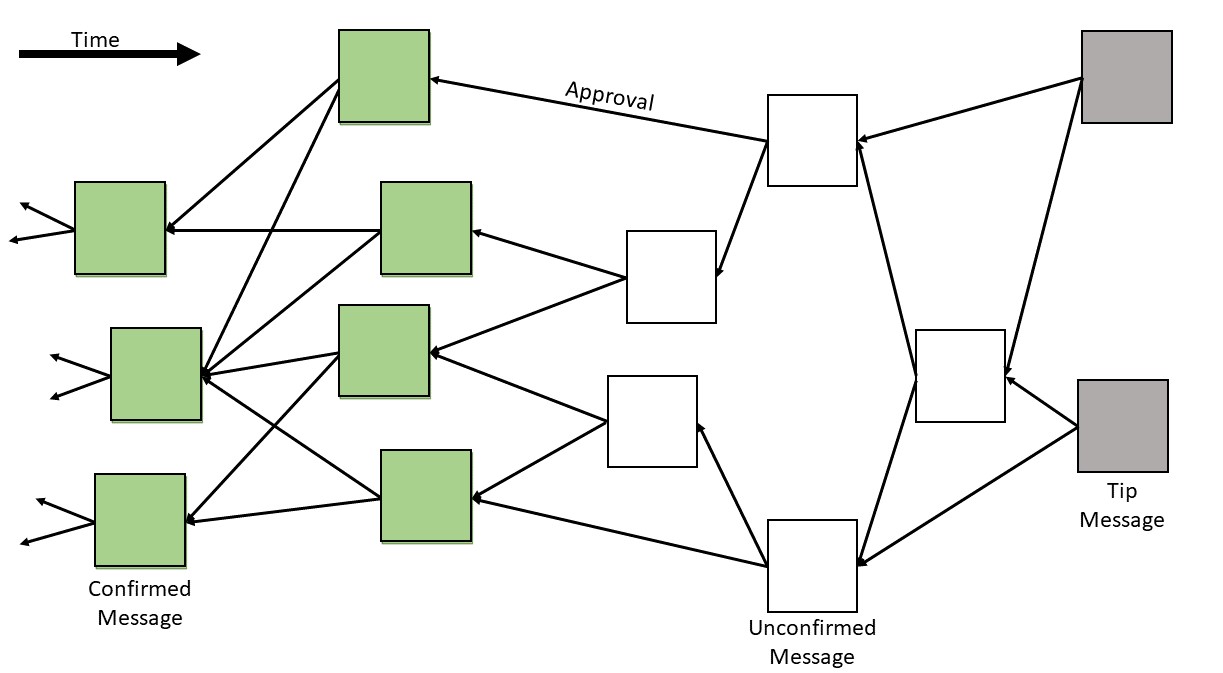}}
    \hfill
      \subfloat[\label{blk}]{%
       \centering
    \includegraphics[width=\linewidth]{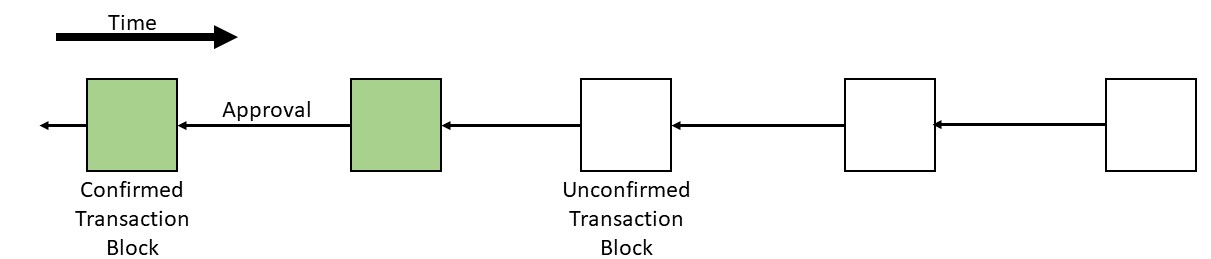}}
    \caption{\ref{dag} shows the Tangle topology; \ref{blk} shows the blockchain topology.}
    \label{fig:dag}
    \vspace{-1.5em}
\end{figure}

IOTA Tangle is a DAG-based DLT, specifically targeting IoT networks and devices. \textcolor{black}{The Tangle network comprises nodes that use P2P communication to propagate messages and transactions. The nodes are entry points for data and transactions into the network and autonomously handle and verify messages and maintain the state of the Tangle (ledger). Each node has its own local view of the ledger and their decentralized and autonomous nature are key to the operation of the Tangle protocol.} The DAG structure of the Tangle ledger, as well as the blockcain topology, is shown in Figure \ref{fig:dag}. \textcolor{black}{When issuing each new message (tip), network nodes must first approve two previous messages.} Unconfirmed messages are considered confirmed once an accepted level of network consensus has been reached. \textcolor{black}{As the number of transactions per second (TPS) increases, the confirmation time decreases due to the requirement of each node having to approve two messages when issuing one.} Consensus aspects are covered in detailed later.  
Numerous works have compared IOTA Tangle to alternative DLTs and it performs well, if not best in class, for many key metrics \cite{Salimitari2018}.

\begin{table*}[ht]
\caption{Key differences between legacy IOTA and IOTA 2.0}
\centering
\begin{tabular}{p{0.25\linewidth}p{0.25\linewidth}p{0.25\linewidth}}
\hline
\toprule
\textbf{Feature} & \textbf{IOTA 2.0} & \textbf{Legacy IOTA}\\
\hline
\midrule
Smart contracts  & Supported & Not supported \\ \hline
Digital asset support  & Yes                                                       & No                                                \\\hline
Transaction size       & 100 bytes                                                 & 1700 bytes                                        \\\hline
Decentralization       & Fully decentralized                                       & Coordinator as point of centralization               \\\hline
Sybil protection       & Mana reputation system                                    & None                                              \\\hline
Spam prevention        & Lightweight Adaptive PoW                                  & PoW                                               \\\hline
Address types          & Reusable                                                  & One-time use                                      \\\hline
Consensus mechanism    & FPC binary voting protocol                                & Weighted MCMC tip selection with coordinator         \\\hline
Scalability            & Very scalable (increased TPS with increased network size) & Limited (coordinator and milestones limit scalability) \\\hline
Approvement finality   & Based on consensus mana approval weight – no orphans      & Based on MCMC weight magnitude – orphans possible\\
\hline
\label{contable}
\end{tabular}
\end{table*}

However, upon first going live in 2016, IOTA could not fully realize their vision \cite{Popov2018}, and instead relied on a bootstrap solution called the \emph{coordinator}. The coordinator added a non-ideal element of centralization into the protocol and since then, the IOTA foundation has worked to change their protocol to remove the coordinator in a development arc called \emph{coordicide} \cite{Popov2020a}. Not only is the coordinator an unwanted element of centralization, it also limits scalability of the protocol. Lack of smart contracts and digital asset support were also the key concerns for industry adoption. 

In 2021 IOTA’s first fully decentralized network was released on their development network \cite{misc7}. The protocol, named IOTA 2.0, is significantly and notably different from the legacy protocol and addresses many of its shortcomings, in regard to IoT applications.
Developed in a modular fashion on IOTA’s development network (devnet), IOTA 2.0 is the future of the protocol. Development work continues but a working implementation, nicknamed \emph{Nectar}, is live, with a network migration to the main network (mainnet) planned. \textcolor{black}{The \emph{Nectar} network can support up to 1000 TPS with confirmation times on average between 10 and 12 seconds.} It is for this reason a switch in focus from the legacy implementation to that of the decentralized network is recommended for industry. 

This paper reviews the key protocol changes in IOTA 2.0 with an emphasis on potential impact and benefits to the IoT ecosystem. \textcolor{black}{Exploring the evolution of IOTA Tangle is important from the viewpoint of new IoT-centric opportunities. By providing a technical overview of IOTA 2.0, this work shows the suitability and purposeful design of different components for IoT networks and applications. It provides simulations-based performance evaluation providing insights into the variables affecting the new consensus algorithm.} It also discusses key directions for further research and development for IOTA.

\section{Overview of IOTA 2.0}
\begin{figure*}
    \centering
      \subfloat[\label{hll}]{%
       \centering
    \includegraphics[scale=0.32]{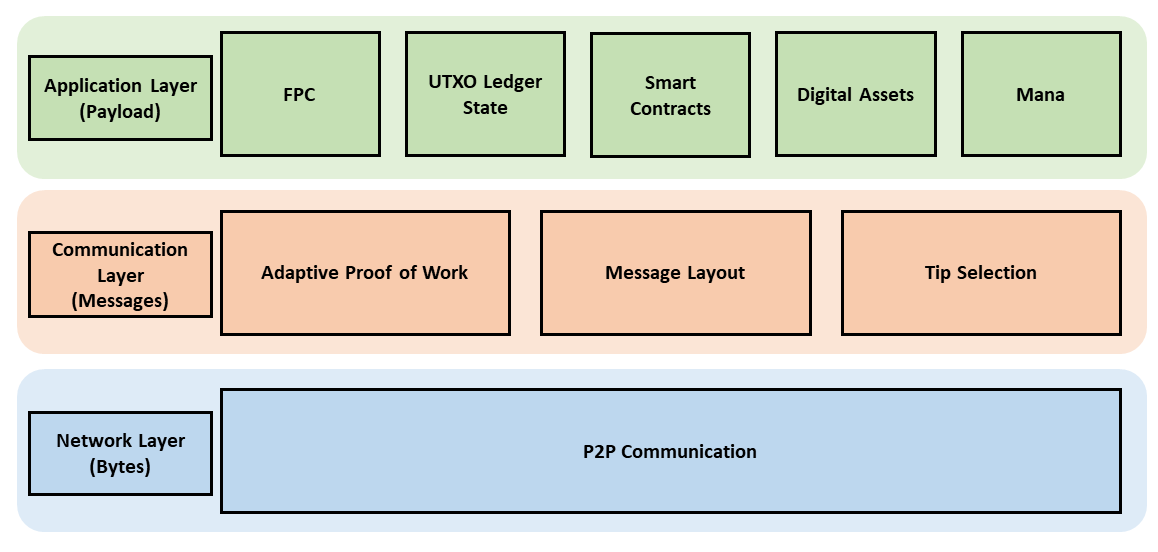}}
    \hfill
      \subfloat[\label{hlo}]{%
       \centering
    \includegraphics[scale=0.45]{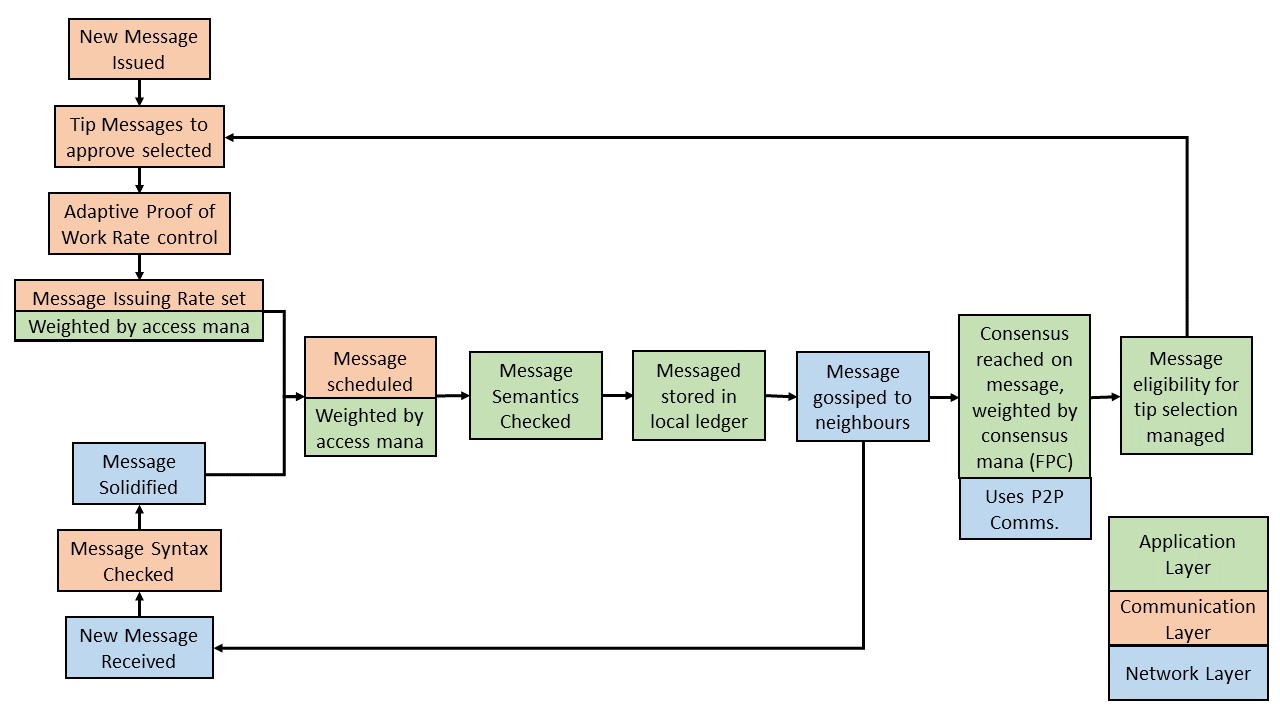}}
    \caption{ \ref{hll} shows abstract protocol layer model for IOTA 2.0; \ref{hlo}  describes high-level flow diagram of IOTA 2.0 protocol. }
    \label{hloll}
    \vspace{-1.5em}
\end{figure*}

The differentiating aspects of IOTA 2.0 can be split into modules \cite{misc6} in order to understand their impact on the overall protocol as well as to IoT  implementations. \textcolor{black}{Security benefits of highlighted features are mentioned in this works but for a full survey on IOTA's security please see \cite{CONTI2022103383}.} Table \ref{contable} summarizes some of the key differences between the legacy IOTA and IOTA 2.0, which are discussed in this section. Figure \ref{hloll} provides a holistic view of the overall structure and protocol operation of IOTA 2.0. 
\textcolor{black}{The network layer handles the Tangle node functionality that works with bytes, and therefore encompasses all the  P2P (inter-node) exchanges. The communication layer handles IOTA messages. This includes the required tip selection approvals, rate control mechanisms,  as well as handling the Tangle ledger itself. The application layer deals with executing message payloads. For example, for a value transaction it executes the required consensus, fund transfers, and reputation (mana) transfer.}

It is emphasized that each of the shown modules is under continuous review and development by the IOTA Foundation. We summarize the functionality of each module while highlighting its benefits to the overall protocol and IoT networks. 

\subsection{Adaptive Proof-of-Work}
IOTA uses an explicit rate control mechanism to protect against spam or denial-of-service (DoS) attacks. In the legacy version this took the form of a proof-of-work (PoW) puzzle of set uniform difficulty that every node must complete before attaching a transaction. This ties some workload weight to the attachment thus limiting the rate at which any one node can attach transactions. However, IOTA 2.0 improves on this by implementing an adaptive PoW \textcolor{black}{\cite{8751358}}. Instead of having a set difficulty, defined by minimum weight magnitude (MWM) in legacy IOTA, the protocol instead adapts the difficulty based on a nodes message rate during a set time interval. The more messages a node tries to issue within said given period, the harder the PoW puzzle becomes, and therefore, the lower the rate at which said node can attach messages until it becomes computationally impossible. This protects against burst transaction, and spam and DoS attacks whilst also making the PoW difficulty lower for honest nodes.

\subsection{Fast Probabilistic Consensus}
Consensus in the legacy Tangle was reliant on the coordinator which brought an unwanted element of centralization to the network. With IOTA 2.0, consensus is instead achieved with a \emph{probabilistic leaderless binary voting protocol} called fast probabilistic consensus (FPC). Consensus is needed to maintain the validity of the Tangle by addressing conflicts such as double spends.  In its simplest reduction, FPC allows a node to update its opinion by querying a set size subset of other nodes on the network and then choosing the majority opinion. This is done multiple times (rounds) until an unchanging opinion is decided or a maximum round threshold is reached. For increased resistance to attackers, opinions are weighted by a node's \emph{mana} (discussed later) and random thresholds are used throughout the protocol to prevent meta-stable situations\cite{Popov2020}. The random thresholds are common amongst the nodes and set via a decentralized random number generator (dRNG) module.

\begin{figure} 
    \centering
  \subfloat[\label{na}]{%
       \centering
  	\includegraphics[scale=0.45]{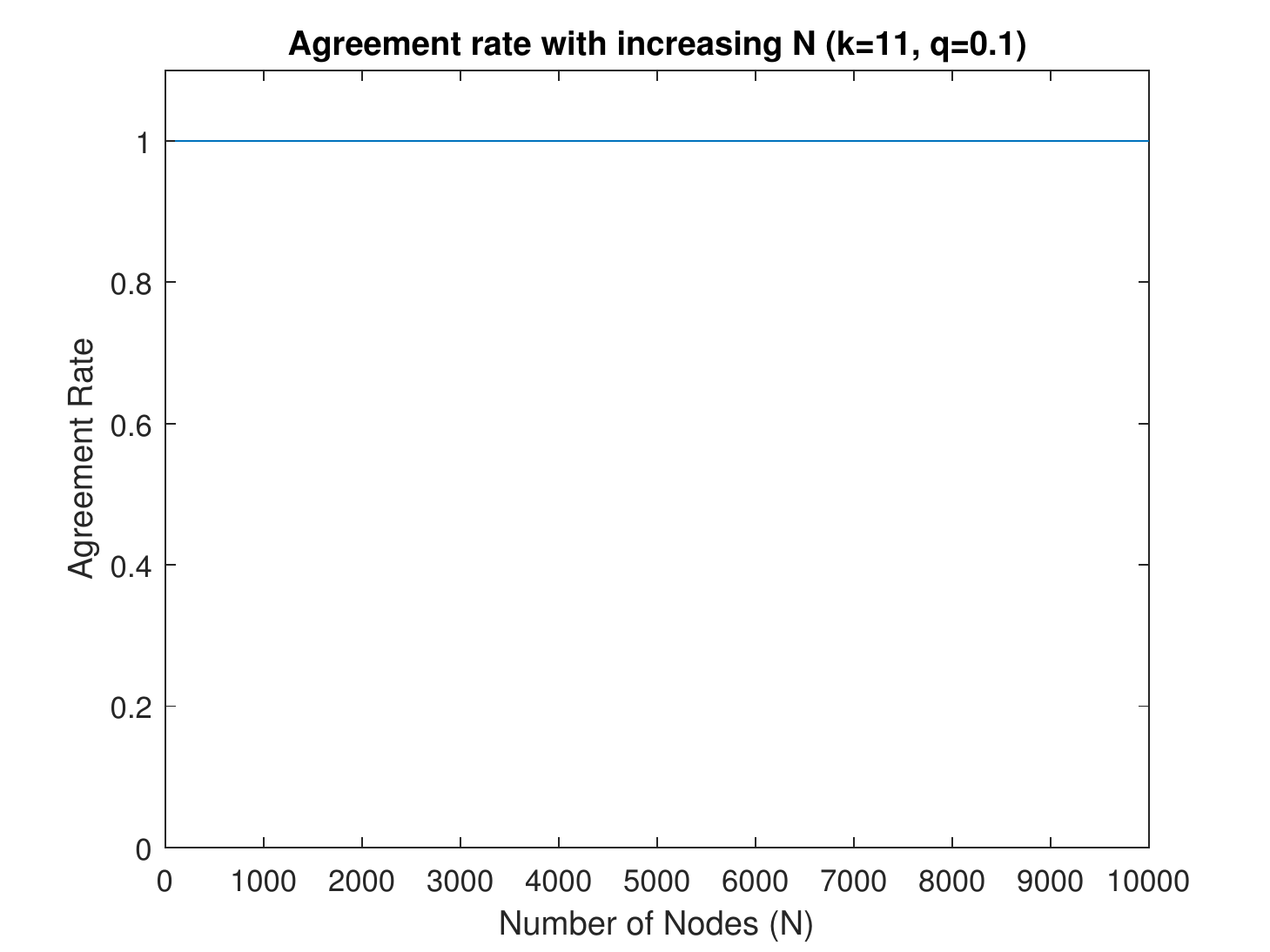}}
    \hfill
      \subfloat[\label{ka}]{%
       \centering
  	\includegraphics[scale=0.45]{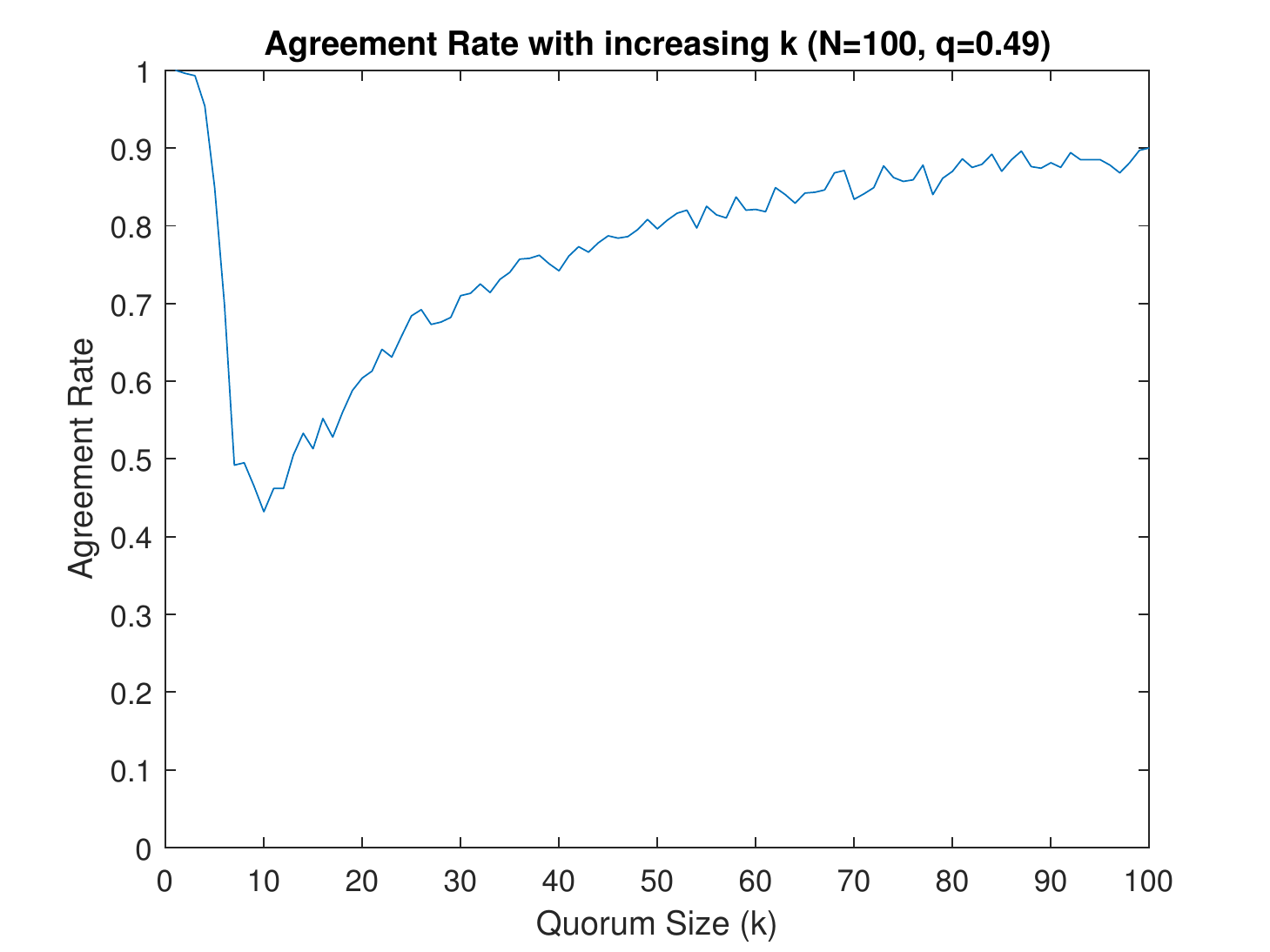}}
     \hfill 
  \subfloat[\label{qa}]{%
       \centering
  	\includegraphics[scale=0.45]{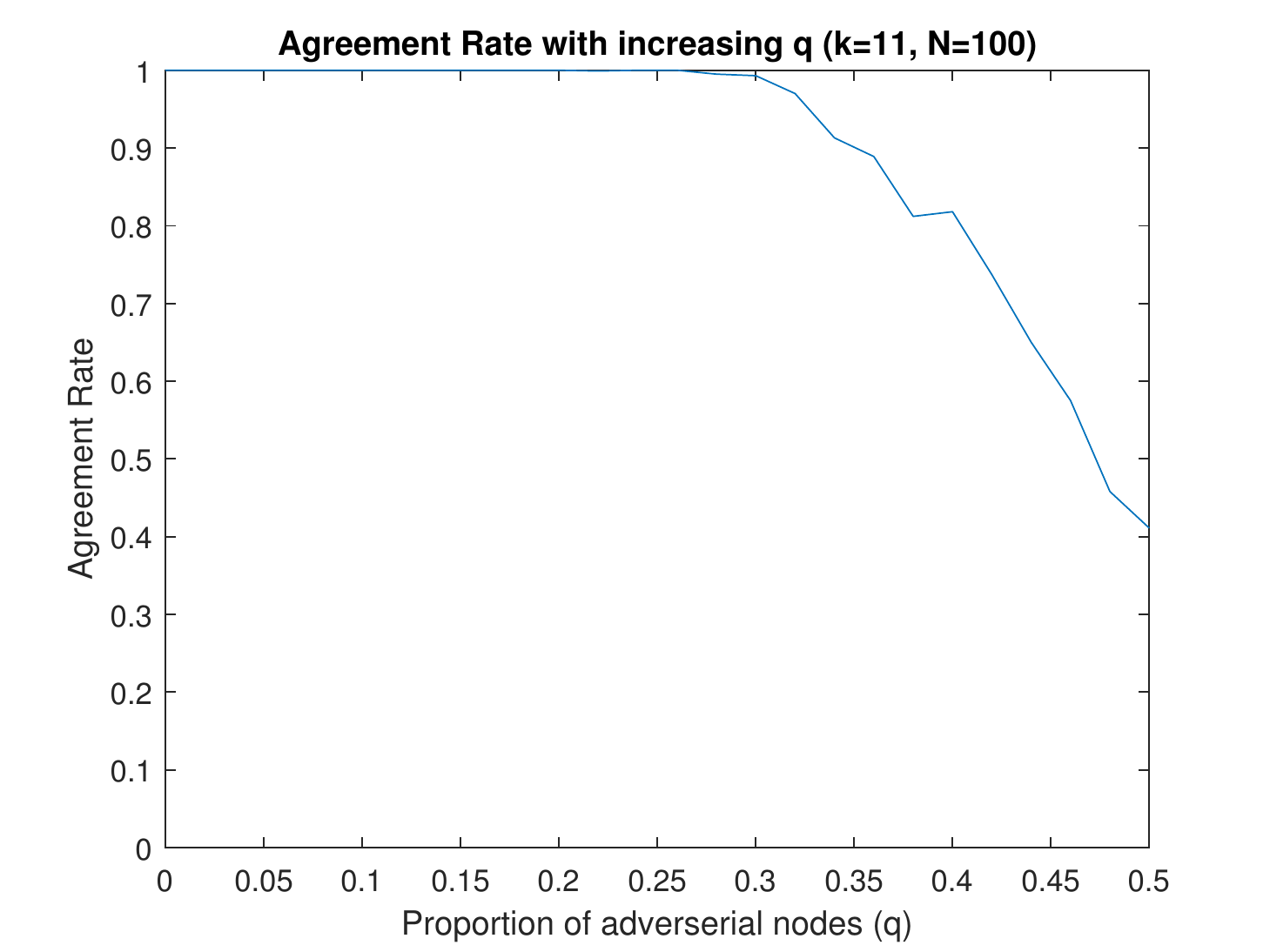} }

	\caption{FPC simulations for agreement rate.}
\label{fig1} 
\vspace{-1.5em}
\end{figure}

\begin{figure}
\centering
  \subfloat[\label{nmt}]{%
        \centering
  	\includegraphics[scale=0.45]{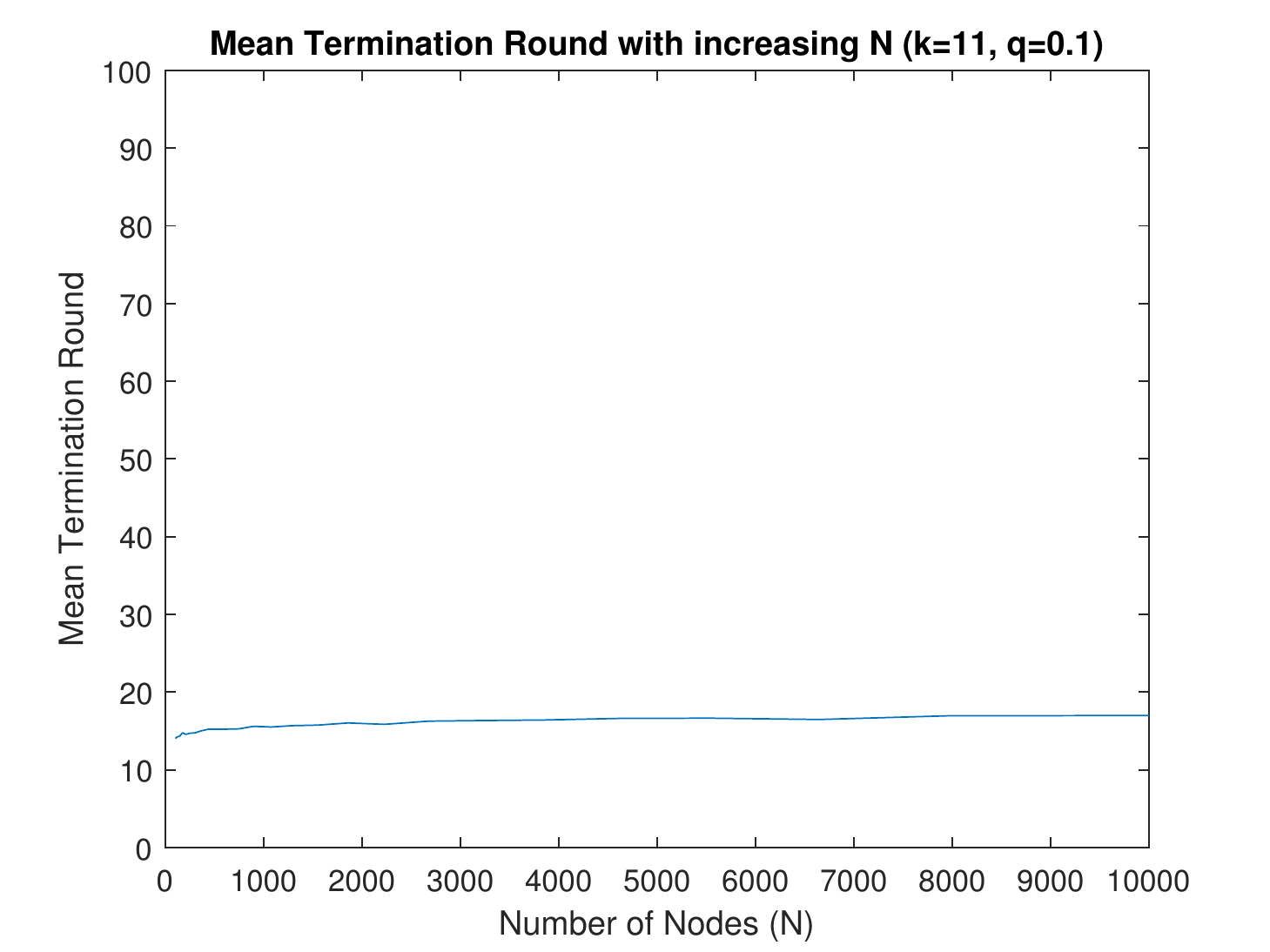} }
    \hfill
  	  \subfloat[\label{kmt}]{%
        \centering
  	\includegraphics[scale=0.45]{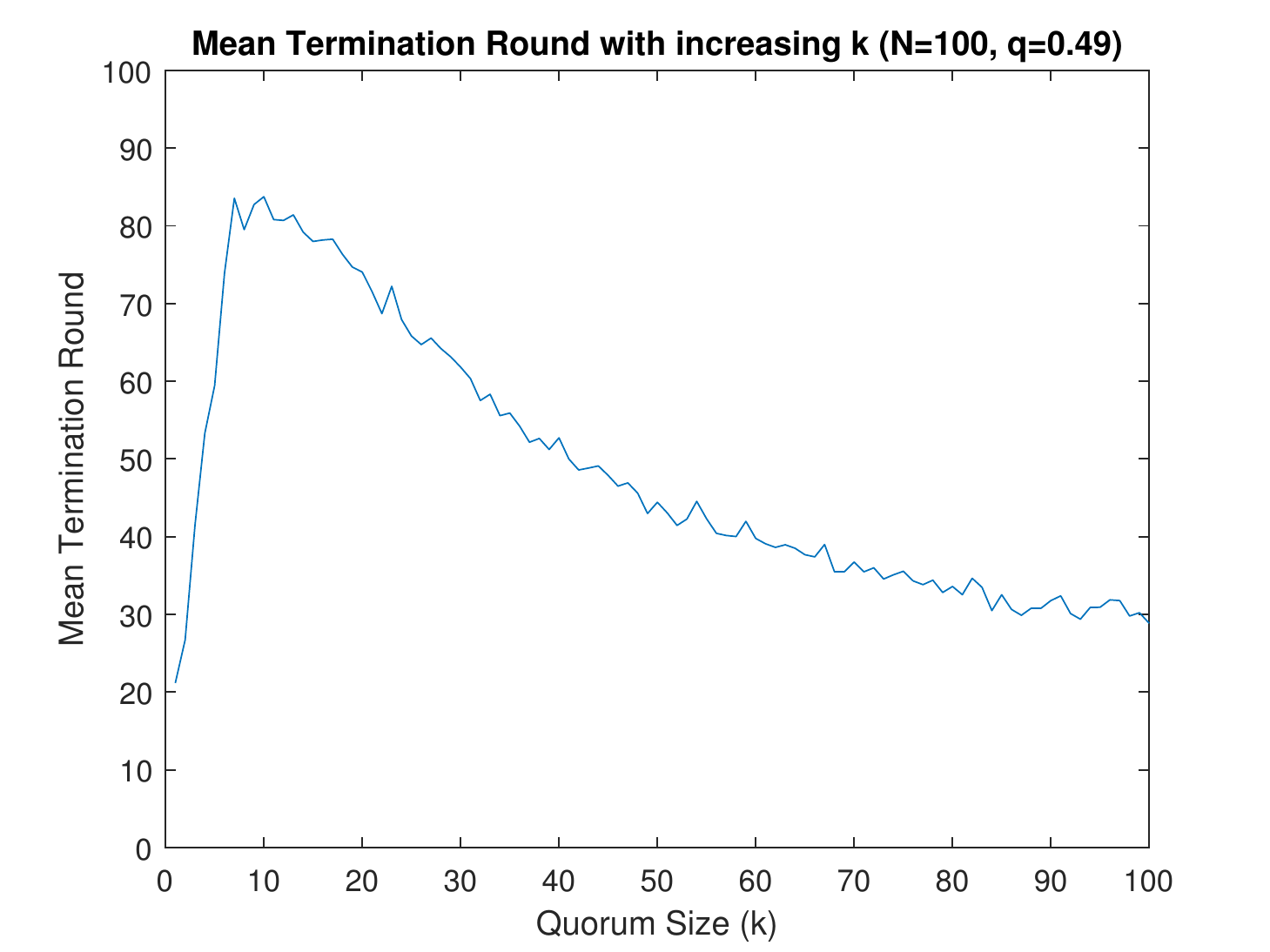} }
  	\hfill
  \subfloat[\label{qmt}]{%
        \centering
  	\includegraphics[scale=0.45]{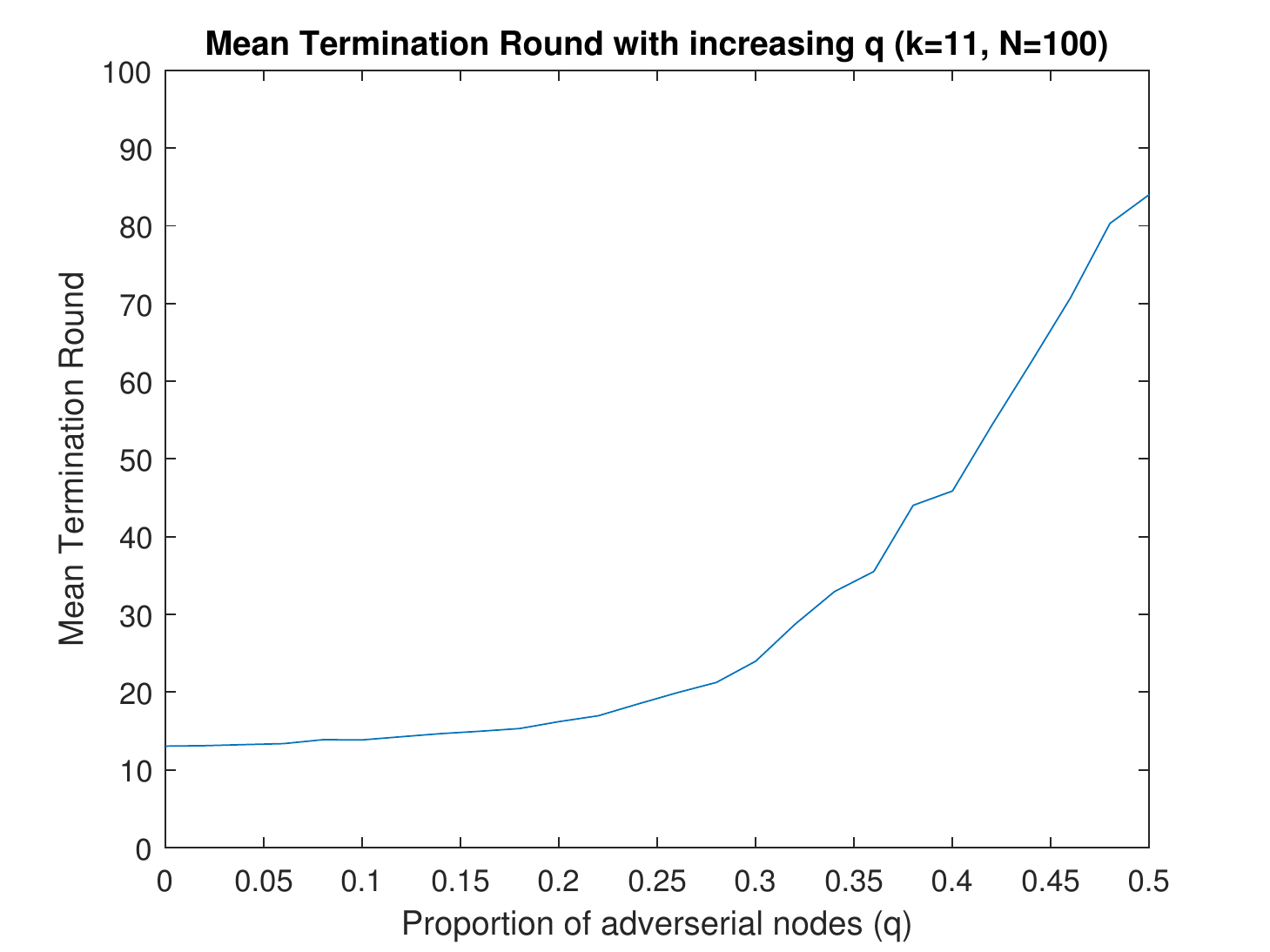} }
  \caption{FPC simulations for mean termination round.}
\label{fig2}
\vspace{-1.5em}
\end{figure}

There are various variables which affect FPC performance, particularly in terms of metrics like \emph{agreement rate} and \emph{mean termination round} of the protocol. We have conducted performance evaluation of FPC using the simulation software by the IOTA Foundation \cite{misc4}. Fig. \ref{fig1} shows the agreement rate of FPC when changing the total number of nodes (N), the quorum size (k) which is the number of neighbors each node queries every round, and the proportion of adversarial nodes in the network (q), respectively. Similarly, Fig. \ref{fig2} shows the corresponding mean termination rounds of the protocol. Fig. \ref{na} and Fig. \ref{nmt} demonstrate the scalability of the protocol with no change in agreement rate and the average finalization round also remaining effectively unchanged, even with increasingly large changes to the number of nodes. As shown in Fig. \ref{ka} and Fig. \ref{kmt}, controllable variables such as the number of nodes queried each round must be adapted to the specific network implementation in order to maximize performance at a manageable communication overhead. It can be seen in Fig. \ref{ka} that if k is too low, most of the nodes won't query an adversarial node, providing a seemingly high agreement rate but masking the adversarial attack. When k is set to the size of the network the highest healthy agreement rate can be seen and the low termination round is observed in Fig. \ref{kmt}; this comes at the cost of a very high communication overhead that some networks could not support. Fig. \ref{qa} and Fig.  \ref{qmt} demonstrate the algorithm's robustness to adversarial nodes up to the unrealistic point where half of the network is maliciously controlled.

\subsection{Tip selection}
\textcolor{black}{With IOTA 2.0 implementing a dedicated FPC voting mechanism, the health of a message and thus its eligibility for tip selection can be decided before any tip selection algorithm} This means as opposed to tip selection playing a vital role in consensus, as in the legacy implementation, its purpose now is only to allow the Tangle ledger to grow in a stable and secure way. This depreciates the value derived from any recent studies that focused on the tip selection algorithm as a component of consensus.

IOTA 2.0 uses restricted uniform random tip selection (RURTS) which with uniform probability selects a tip from a list of tips defined as eligible. The number of approvals a message carries out is now also variable from the original two up to a maximum of eight. The higher number of approvals is used in times of congestion to prevent a spam attack from widening the Tangle and therefore maintaining the healthy growth of the ledger.

\subsection{Mana}
Mana can be seen as a new reputation system for nodes with the primary objective of securing the network against Sybil and Eclipse attacks \cite{misc5}. Each node holds an amount of decaying mana with this quantity being stored and recorded as an extension to the Tangle ledger. When a value transaction is processed, a node ID to ‘pledge’ access mana to is specified causing an increase to that node’s access mana and thus reputation in the network. When a node takes part in FPC, it is ‘pledged’ consensus mana. The combination of access mana and consensus mana constitutes a nodes total mana. Aside from processing value transactions, nodes can gain mana by holding tokens (larger token balance the larger mana gained) or renting mana from a higher mana node. A node's mana decays over time.

Mana protects against Sybil attacks across the protocol with key applications in four modules. Access mana is used in congestion control to limit the data a node can attach to the Tangle as such that the data any one node can add is proportional to their share of the active access mana in the network. The probability a node is queried in FPC, as well as to be selected during dRNG, is also proportional to the amount of consensus mana a node holds. The opinion of a node during FPC is similarly weighted by its consensus mana. \textcolor{black}{The use of mana during consensus, and the difficulty in acquiring mana, is key to defending Sybil attacks. New malicious identities are therefore easily distinguished by nodes trusted by the network by their amount of held mana.} The most influential nodes (high mana nodes) are also protected against Eclipse attacks as during auto peering nodes peer with others with a similar amount of mana.

\subsection{Unspent Transaction Output Model Ledger}
Due to other changes in the protocol, a need is established to both easily and quickly detect double spends and transactions that try to spend non-existing funds and determine a node's initial opinion for every received transaction. To facilitate this, the ledger model is changed to that of an unspent transaction output (UTXO) model which enables real-time transaction payload validation.

Unlike the legacy implementation, the UTXO ledger’s balances are not directly associated with addresses and instead are tied to the output of transactions. Inputs to new transactions are then defined via the output of old transactions. This means funds are individually specifiable and the same addresses can now be reused with multiple transactions without a loss in security. Double spends are addressed via the \emph{reality based ledger state} which models the possible realities of the ledger state thus identifying double spends and unmergeable branches and instead taking the reality that offers the most healthy Tangle end state. This information is all available without having to walk the Tangle \cite{misc6}.

\subsection{Message layout}
The data carrying objects of the IOTA 2.0 protocol are called \emph{messages} as opposed to the previously employed \emph{transactions}. Messages are split into three parts: the header, the payload and the signature. The header includes the IOTA version number, message parents, issuing timestamp, issuing node ID, and a proof-of-work (PoW) nonce. The payload stores the intended content of the message and can be of many types whether that be plain data, a value transaction, or a user-defined custom payload. Depending on the payload type, the message is parsed in a different way with any application being able to define a payload and thus how it interacts with the Tangle. The message ends with a signature from the issuing node which signs the entire message, including the payload, making the entire message unalterable.


\subsection{Smart contracts}
Lack of support for smart contracts has been a barrier in widespread industry adoption of IOTA Tangle. IOTA smart chain protocol (ISCP) implements smart contract functionality to the new decentralized network and does so in a way that is different to current alternatives like Ethereum \cite{iotasm}.

ISCP uses off-chain smart contracts, as to say that the contracts are not executed and secured by the entirety of the main network. Instead, smart contracts are run on sub-chains connected to the main Tangle and maintained by a subset of nodes called the committee. The committee handles the consensus and execution of all contracts on its chain and updates the main Tangle, and thus the rest of the network, as to their state with signed transactions. This means transaction costs are very low and can be more easily forecast. ISCP operation does not contribute to the operational stress of the whole main network and the security of any smart contract is variable and can be increased by increasing its committee size. 

\textcolor{black}{Although IOTA smart contracts are written in their own language, the ISCP supports an Ethereum virtual machine. This allows smart contracts written in solidity for the Ethereum DLT to be run on an IOTA Tangle network. This facilitates the reuse of exiting Ethereum contracts and allows them to leverage the benefits of the IOTA network such as feeless transactions and a faster execution speed.}

Smart contracts are configured and deployed by their owner but are run and maintained by the committee. The owner can set a variable reward to incentivize committee nodes to process their contract. Fees are again flexible and dependant on the chain, contract and its owner.

\subsection{Digital Assets}
Digital assets have been a promising feature of many other DLTs with the legacy IOTA not supporting them. However, with the changes to the protocol bought about by IOTA 2.0, IOTA will now support digital assets with the main short falling of other DLTs’ implementations addressed. Due to the feeless transactions of IOTA, they will support the creation of digital assets that are secured by the main Tangle, without any fees. This will allow the free digital twinning or tokenization of any system or asset; physical or otherwise \cite{misc1}.

\section{IOTA 2.0 for IoT}
This section discusses the key benefits of each module change in relation to IoT networks and applications. 

\subsection{Consensus Enhancements for IoT}


The implementation of a dedicated consensus module and associated algorithm into the IOTA protocol is key for achieving full decentralization, and therefore of importance to IoT network implementations. As a consensus protocol, FPC is energy-efficient and lightweight, doesn’t require staking, and is highly scalable; making it attractive to IoT networks. The communication overhead required for direct queries of FPC can be of concern for some IoT implementations; however, techniques to overcome this are discussed later.

With adaptive PoW, nodes are rewarded for honest behaviour with the computational intensity of the puzzle being solved remaining low. By making this process more lightweight than the constant difficulty PoW previously used, it is of benefit to resource constrained IoT devices.

\subsection{Tip Selection Enhancements for IoT}
The benefits of the changes to tip selection for IoT networks are that tip selection becomes both substantially faster and more lightweight. The previously employed weighted Markov Chain Monte Carlo (MCMC) algorithm required walking the Tangle and the updating of weights, both of  which had an associated computational cost. These elements are not required by RURTS with the time complexity moving from the $\mathcal{O}(n^2)$ of MCMC to $\mathcal{O}(n)$ for RURTS. There is also an increased reliability in the transactions in the network and a significant reduction in orphaned messages and the resulting  need for re-attachments.

This has obvious benefits when applied to IoT networks. For honest nodes, the rate setting PoW becomes less computationally intensive which is good for lightweight IoT devices. This change also facilitates a higher healthy message attachment rate as healthy behaviour is less limited by the PoW as the difficulty can be assumed to be low. 

\subsection{IoT Security Benefits via Mana}
Mana brings clear security benefits to IoT networks looking to utilize IOTA with the threat of an attacker gaining increased control over the network by creating multiple identities being nullified. Mana offers a higher level of protection to this kind of attack than the previously employed tip selection algorithms whilst also more astutely tackling congestion control and increasing the robustness of the employed consensus protocol.

\subsection{IoT-centric Implementation Enhancements}
The change of the ledger state, as discussed previously, helps facilitate some of the other module changes in the protocol. Its implementation provides necessary information previously not available without walking the Tangle. This is a benefit for IoT networks as it avoids the associated computational and time cost of both walking the past cone of a message as well as solidification that would have been required to do so.

The changes to the message layout, as discussed previously, allow for applications to build message payload types specifically designed to their use case, which is attractive for any network looking to employ IOTA. The change to atomic messages also provides a massive reduction in the transaction size used by legacy IOTA. Previous transactions were not size variable and were fixed at around 1.7 kb with atomic transactions being as small as 100 bytes \cite{misc3}. This makes them much more suitable for lightweight IoT devices and reduces communication overhead. 

\subsection{Smart Contracts for IoT}
The ability to use feeless smart contracts as well as those incentivized with micro-transactions is extremely attractive to IoT networks of which many were not resource rich enough to participate in Ethereum smart contracts.

The key differences between the popular alternative Ethereum, and IOTA are summarized in Table \ref{sctable}. Clear benefits to IoT networks are shown, such as reduced transaction fees and reduced network load. However, security must be considered on a individual contract basis so the necessary degree of decentralization and security can be picked via the committee size. Contracts which hold a large value for example, should have an appropriately large committee size and thus level of security.

\begin{table*}[ht]
\caption{Key differences between Ethereum and IOTA smart contract protocols}
\centering
\begin{tabular}{p{0.25\linewidth}p{0.25\linewidth}p{0.25\linewidth}}
\hline
\toprule
\textbf{Feature} & \textbf{IOTA Smart Contracts} & \textbf{Ethereum Smart Contracts}\\
\hline
\midrule
Transaction   fees                              & Very low   (feeless), supports microtransactions, forecastable                                       & Higher   transaction fees + gas, no micro-transactions                                 \\
\hline
Network   utilization                           & Uses a subset   committee of nodes of variable size                                                  & Each contract   executed by the whole network                                          \\
\hline
Security                                        & Variable   (dependent on committee size)                                                               & Set and   scales with network size – always high                                       \\
\hline
Layer                                           & 2 – off-Chain                                                                                  & 1 – on-chain                                                                     \\
\hline
Speed                                           & Parallel   transactions, parallel smart contract chains, only committee nodes involved, very fast  execution & No parallel transactions,   whole network secures each contract, slower execution time \\
\hline
Scalability                                     & High (smart contracts executed   in parallel by various committees and chains)                          & Poor (each contract   executed by whole network)                                \\
\hline
\label{sctable}
\end{tabular}
\end{table*}

\subsection{Asset Digitalization for IoT}
Digital assets, like non-fungible tokens (NFTs), allow for anything from art, IoT devices, or industrial machinery to be digitally represented, providing proof of both authentication and ownership. Attractive applications of digital assets to an industrial and IoT setting become apparent when you also consider IOTA’s support of both encrypted, immutable, and verifiable data as well as feeless transactions. These in combination could facilitate the digital twinning of a whole system with physical assets being digitally tokenized and all their associated data flows being handled and mapped by the Tangle. All of this whilst being secured by the Tangle. IOTA believe this will \emph{lead to the creation of the machine economy} \cite{misc1}.

\section{Directions of Further Research}
The development network implementation of IOTA 2.0 is under constant review. All modules are being fine-tuned and edited based on the analysis of their operation on the IOTA Foundations development network implementation. This section aims to cover research directions and areas of development that will have sizeable impact on IoT Tangle implementations and thus the wider IoT ecosystem. These changes aim to improve scalability, reduce confirmation times, and improve data credibility amongst others. Three main development directions are discussed within this section: \textcolor{black}{consensus, oracles, and sharding}.

\subsection{Consensus - On Tangle FPC}
IOTA continues to develop and research improvements to its consensus protocol. The most recent development direction focuses on on-Tangle fast probabilistic consensus (OTFPC) \cite{rs}. This takes some elements of the previously described FPC and combines them with a virtual voting protocol that considers approval weight (AW), which is dependent on the number of indirect and direct approvals of the message, in order to provide finality based on the heaviest branch in double spend scenarios. More formally, AW is `the percent of the active consensus mana of nodes who directly or indirectly reference it' \cite{misc8}. A double spend conflict is resolved when one of the conflicting transactions reaches a threshold AW and thus its branch is finalized. To minimize the amount of orphaned transactions from the losing branch, only the double spend message's payload is rejected. This allows the subsequent non-conflicting approving messages in the branch to merge with the main Tangle avoiding being orphaned.

OTFPC is also run in rounds but a key difference is that instead of asking nodes to vote on their opinion via direct queries, a node simply referencing that message is considered equivalent to a positive vote as nodes only reference messages they approve of. By tracking the references, and thus approvals, and calculating an approval weight, conflicts can be simply resolved by picking the message or branch with the highest AW. This removes the need for P2P communication between nodes during consensus, reducing the communication overhead, a positive for IoT networks. As all the information needed for consensus is available on the Tangle, confirmation times are also reduced. 

\subsection{From Source Data Issuing - Oracles}
Oracles are a key progression with high relevance to the IoT ecosystem. Oracles facilitate data being uploaded to the Tangle directly from the source, for example IoT sensor devices. This increases the level of trust in the data and removes opportunities for data manipulation which could bring the data’s credibility into repute as there is no intermediary between the data issuer and the attachment of its data to the Tangle. Therefore, lightweight IoT data sources will be able to feed data directly into smart contracts or, as data messages, or streams, stored and secured by the Tangle ledger without having to contribute to the PoW and other protocol processes. 

Further developments include the use of oracle distributed networks (ODNs)\cite{rs}. By combining several oracles, which all provide data relevant to each other, as well as an ‘aggregator oracle’ which connects to the main Tangle ledger, ODNs aim to not only provide direct data issuing but also create a network that enhances the credibility of data by acknowledging and disregarding erroneous data submissions as part of data validation, regardless of data type. Developments as to an aggregator oracle with a built in `truth finding algorithm' aim to consolidate data, checking for and eliminating inconsistencies, before issuing only data deemed correct/truthful to the main tangle. 

Data provider sources and use cases also vary with them not limited to IoT sensors. For example, an oracle could be set to regularly pull data from a publicly accessible feed of a website or even provide potential DLT interoperability by using data from a secondary DLT platform.

\subsection{Enhancing Scalability - Sharding}
Sharding is a technique being researched and developed to further increase the scalability of IOTA Tangle networks \cite{rs}. IOTA Tangle aims to be able to increase its throughput to a level that can facilitate the relevant demands of the whole IoT ecosystem and sharding is the leading development direction to attain this goal.

In its basics, a shard is a subset of network nodes with those nodes processing and sending their messages amongst their subset and not having to process other shards messages. This breaks up the total number of messages nodes must process, reduces redundancy, and with each of these shard networks operating in parallel, the overall network throughput can be increased. A shard can be permissioned with approval needed for nodes to join and control of the sybil protection mechanism and congestion control amongst the shard given. This is of particular interest to IoT use cases that were previously more suited to private implementations; by instead using a permissioned shard running on the main public Tangle network the privacy benefits of a permissioned setup are given with the added benefits brought by connection to the public network.
The IOTA Foundation data sharding approach is in development and plans to use a hierarchical sharding approach to increase overall network throughput for data messages. Shards update the main tangle as to messages they have processed via stamp messages to their parent shard with consensus being reached upon said stamp and being necessary for confirmation of the original message. Sharding can be multiple levels deep with stamping occurring recursively to the parent shard until the main Tangle network is stamped. This ensures a proof of inclusion for any message in any shard and allows the main tangle to contribute to the confirmation of all messages no matter the shard depth. 

Other non-hierarchical sharding techniques are in development with fluid sharding being an active research direction that aims to infinitely increase the scalability of the network and create a network that can optimally support any device regardless of individual device requirements. Early research suggests an approach that sees all nodes as individual shards with interaction occurring with other shards within a radius of perception. 

\section{Concluding Remarks}
The key objective of removing the coordinator is vital to the wide adoption of IOTA Tangle into the IoT ecosystem. Removing the coordinator achieves a greater degree of scalability for IOTA 2.0 as well as other benefits of full decentralization such as removing the single point of failure and a necessary trust in a central authority. Other changes, such as mana, offer improved Sybil protection and new features, like smart contracts and digital assets, show improved functionality when compared to current industry adoption leaders. Consensus is also improved upon and is now both more scalable and faster than before. With the reduced need to walk the Tangle and adaptive PoW, the protocol becomes more lightweight for honest nodes increasing its accessibility within the IoT domain. The communication overhead is also reduced with the use of atomic messages.

With the ongoing developments discussed, the IOTA Tangle looks to become increasingly more scalable, with a smaller communication overhead and provide the means of direct from source data attachment of integrity verified data. These developments continue to position IOTA Tangle as a highly attractive DLT for IoT industry.


%



\bibliographystyle{IEEEtran}
\bibliography{references.bib}
\end{document}